\documentstyle[pre,aps]{revtex}

\begin{document}

\draft
\title{Density fluctuations and entropy}

\author{J.A. Hernando\cite{jah}}
\address{Dept. of Phys., Comision Nacional de Energia
Atomica,\\Av. del Libertador 8250, 1429 Buenos Aires, Argentina}

\author{L. Blum\cite{lb}}
\address{Dept. of Phys., POB 23343, Univ. of Puerto Rico,\\
 Rio Piedras, PR 00931-3343, USA}

\date{\today}
\maketitle

\begin{abstract}
A new functional for the entropy that is asymptotically correct both in the
high and low density limits is proposed. The new form is
\[
S=S^{(id)}+S^{(ln)}+S^{(r)}+S^{(c)}
\]
where the new term $S^{(c)}$ depends on the p-bodies density 
fluctuations $\alpha_p$ and has the form
\[
S^{(c)}=\left\langle N\right\rangle \left\{ \ln 2-1+\sum_{p=2}^\infty
\frac{\left( \ln 2\right) ^p}{p!}\alpha _p-\left[ \exp (\alpha _2-1)-\alpha
_2\right] \right\} +\hat S\nonumber
\]
$\hat S$ renormalizes the ring approximation $S^{(r)}$. This result is obtained by
analyzing the functional dependence of the most general expression of the
entropy: Two main results for $S^{(c)}$ are proven: i) In the
thermodynamic limit, only the functional dependence on the one body
distribution function survives and ii) by summing to infinite order the
leading contributions in the density a new numerical expression for the
entropy is proposed with a new renormalized ring approximation included. The
relationship of these results to the incompressible approximation to entropy
is discussed.
\end{abstract}

\pacs{05.20.-y, 65.50.+m}


\section{ Introduction}

Entropy is one of the very important and challenging thermodynamic
quantities in statistical mechanics; this is so because it depends on all
the n-particle distribution functions. The problem is not the lack of exact
expressions but to derive equations that are both accurate and manageable
from a theoretical and numerical point of view. Amongst the exact
expressions we cite the classical work of Nettleton and Green \cite{ngreen},
and, more recently, \cite{jah1,puos}, while approximate expressions
can be found in \cite{puos,jah2,bush,bev1,wall1}. In \cite
{jah2} we have shown how an infinite subset of terms (dependent exclusively
on the one- and two- body distribution functions) can be analytically summed
giving rise to the so-called ring approximation and, through a minimization
of the free energy functional, the well known HNC\ approximation \cite{mohi}
is obtained as an optimized superposition approximation. Later on, Bush 
et al \cite{bush} derived sets of integral equations by analyzing several
levels of approximation to the grand potential function and, in a recent 
and very interesting article, Puoskari \cite{puos} extends the ring
approximation to three particle functions showing how the HNC2 equations
either of the Wertheim \cite{wer1} or the Baxter \cite{bax} variety can be
obtained. Baranyai and Evans \cite{bev1} showed that, even though the
derivations are done in the grand canonical ensemble, the entropy equations
are, in fact, ensemble invariants if local expressions are used for the
entropy. In this way, the comparison with canonical ensemble numerical
simulations is justified. They also analyze the convergence range of some
needed integrals showing how this range increases at high densities. Wallace
\cite{wall1} also worked with the same type of expression and, by analyzing
the behaviour of density fluctuations, proposed an incompressible
approximation for the entropy of dense fluids (which, as Wallace himself
states, is decidedly wrong in the low density regime). He then concluded
that the dilute gas and dense liquid regimes occupy different regions of the
phase space. These conclusions have been criticized \cite{bev1}. Laird and
Haymet \cite{lhay1} have extended the ring approximation to mixtures and
applied it to electrolytes. They found that the correct Debye-Huckel
expression for the entropy in the low concentration limit is obtained when
the ring approximation is included. They have also discussed \cite{lhay2}
the incompressible approximation in dense fluids, proposed an expression
that differs from the one of Wallace, and applied it with good numerical
results.

Summarizing, these works show that, at low densities the inclusion of the
ring approximation gives a very accurate entropy equation and the
incompressible approximation is hopelessly wrong, while, at high densities,
the ring term overestimates the entropy and the incompressible approximation
is reasonably accurate. It has been suggested that more than two bodies
correlations must be incorporated in order to have an accurate expansion
and, in this respect, Puoskari's work \cite{puos} is quite promising. It is
the purpose of this article to elaborate on the compressibility related
contribution to entropy and improve the entropy expansion when truncated to
the pair distribution level. In Section II we discuss the conceptual
structure of the entropy when written as a functional of the n-bodies
distribution function. In Section III it is shown by functional
differentiation that the compressibility related contribution only depends
on the one body distribution function as well as on thermodynamic parameters
and, in Section IV, by summing three subsets of terms, we shed some light on
the nature of the incompressible approximation and show how the correct low
density limit can also be obtained. In Section V\ we present our conclusions
and propose a new entropy equation.

\section{ENTROPY STRUCTURE}

We can think on two different criteria for the analysis of the entropy. The
first one can be called the functional criterium; is the one we take when
interested in the entropy dependence on distribution functions, e.g., when a
variational principle is formulated. The second point of view is the
numerical one, i.e. the main goal is numerical accuracy. It is clear that it
is not necessarily true that the same expression will fulfill satisfactorily
both goals. In particular, the incompressible approximation (see eq. (\ref
{comp-ap} )) is numerically correct at high densities, but has the wrong
behaviour at low densities and also lacks a sound theoretical foundation.

When the entropy dependence on all the n-bodies distribution functions is
explicitly written, we obtain an approximate expression with the following
distinct structure contributions \cite{puos,jah2}:

\noindent i) The ideal gas contribution

\begin{equation}
\frac{S^{(id)}}k=\left\langle N\right\rangle \left[ \frac 52-\ln (\rho
\lambda ^3)\right]  \label{s-id}
\end{equation}

\noindent where $\rho $ is the number density and $\lambda $ the thermal
wavelength.

\noindent ii) The ever present logarithmic contribution

\begin{equation}
\frac{S^{(\ln )}}k=-\sum_{p\geq 1}\int d\{{\bf p}\}n_p(\{{\bf p}\})\omega
_p(\{{\bf p}\})  \label{s-ln}
\end{equation}

\noindent where $n_p(\{{\bf p}\})$ is the p-particles distribution function
and $\omega _p(\{{\bf p}\})$ the irreducible p-bodies contribution to the
potential of average force. More specifically, the link with the more usual
notation is

\begin{equation}
g_p(\{{\bf p}\})=\frac{n_p(\{{\bf p}\})}{\prod_in_1({\bf i})}  \label{g-def}
\end{equation}

\begin{equation}
e^{\omega _p(\{{\bf p}\})}=\frac{n_p\prod_{\{\bf{p-2\}}\subset
\{\bf p\}}n_{p-2}\prod_{\{\bf{p-4}\}\subset \{\bf p\}}n_{p-4}\ldots
}{\prod_{\{\bf{p-1}\}%
\subset \{p\}}n_{p-1}\prod_{\{\bf{p-3}\}\subset \{\bf p\}}n_{p-3}\ldots }
\label{w-def}
\end{equation}

\noindent As usual, we have that $g_2=1+h_2$ and, through the use of the
generalized superposition approximation (GSA) \cite{fkop} we can write that,
e.g.

\begin{eqnarray}
g_3(\{{\bf 3}\}) &=&\left[ 1+\Delta _3(\{{\bf 3}\})\right] \prod_{\{{\bf 2}%
\}\subset \{{\bf 3}\}}g_2(\{{\bf 2}\})  \nonumber \\
g_4(\{{\bf 4}\}) &=&\left[ 1+\Delta _4(\{{\bf 4}\})\right] \frac{\prod_{\{%
{\bf 3}\}\subset \{{\bf 4}\}}g_3(\{{\bf 3}\})}{\prod_{\{{\bf 2}\}\subset \{%
{\bf 4}\}}g_2(\{{\bf 2}\})}=  \label{delf} \\
&&\left[ 1+\Delta _4(\{{\bf 4}\})\right] \prod_{\{{\bf 3}\}\subset \{{\bf 4}%
\}}\left[ 1+\Delta _3(\{{\bf 3}\})\right] \prod_{\{{\bf 2}\}\subset \{{\bf 4}%
\}}g_2(\{{\bf 2}\})  \nonumber
\end{eqnarray}

\noindent which introduces the family of $\Delta _p$ functions that, when
different from zero, correct for the difference with the GSA. They can also
be written as

\begin{equation}
\Delta _p(\{{\bf p}\})=e^{\omega _p(\{{\bf p}\})}-1  \label{delp}
\end{equation}

\noindent This $S^{(\ln )}$ contribution is the one that, when functionally
differentiated, gives rise to the $\ln g$ contribution in the integral
equations

\noindent iii) The ring term, which in its simplest, two-body version, is

\begin{equation}
\frac{S^{(r)}}k=\frac 12\sum_{p\geq 3}\frac{(-1)^{p-1}}p\int d\{{\bf p}%
\}\prod_{i=1}^pn_1({\bf i})h_2({\bf 12})h_2({\bf 23})\ldots h_2({\bf p1})
\label{s-ring}
\end{equation}

\noindent and it can be summed in homogeneous systems \cite{jah2}. This term
is the responsible of the contribution $h_2-c_2$ in the integral equations.
The three bodies version is derived in \cite{puos}.

\noindent iv) The compressibility related contribution

\begin{equation}
\frac{S^{(c)}}k=\sum_{p\geq 2}\frac 1{p!}\int d\{{\bf p}\}\prod_{i=1}^pn_1(%
{\bf i})\Delta _p(\{{\bf p}\})\Gamma _p(\{n_p\})  \label{s-compa}
\end{equation}

\begin{equation}
\Gamma _p(\{n_p\})=\frac{\prod_{\{{\bf p-1}\}\subset \{{\bf p}%
\}}g_{p-1}\prod_{\{{\bf p-3}\}\subset \{{\bf p}\}}g_{p-3}\ldots }{\prod_{\{%
{\bf p-2}\}\subset \{{\bf p}\}}g_{p-2}\prod_{\{{\bf p-4}\}\subset \{{\bf p}%
\}}g_{p-4}\ldots }  \label{s-compb}
\end{equation}

\noindent Its first term is essentially the compressibility ($\Delta
_{2}\equiv h_{2})$ and the sequence of products in eq. (\ref{s-compb}) stops
when reaching either $g_{3}$ or $g_{2}$ . As far as we know there are no
previous studies of the whole series given in eq. (\ref{s-compa}); the
compressibility approximation focuses on the first term of this series,
which for a one component homogeneous system is

\begin{equation}
\frac{S_2^{(c)}}k=\left\langle N\right\rangle \frac \rho 2\int d{\bf r}%
h_2(r)=\frac{\left\langle N\right\rangle }2\left( -1+\alpha _2\right)
\label{comp-ap}
\end{equation}

\noindent As in the dense liquid limit is $\alpha _2\ll 1$, the
incompressible approximation considers $\alpha _2=0$ in the whole $\rho -T$
space.

\section{FUNCTIONAL DEPENDENCE}

Here we prove that, in the thermodynamic limit, all the functional
derivatives of the compressibility contribution with respect to the
distribution functions can be summarized in the equation

\begin{equation}
\frac{\delta S^{(c)}/k}{\delta n_p(\{{\bf p}\})}=-\delta _{1p}+\bigcirc
(e^{-<N>})  \label{ds-com}
\end{equation}

\noindent Therefore,

\begin{equation}
\frac{S^{(c)}}k=-\int d\{{\bf 1}\}n_1(\{{\bf 1}\})+C(\rho ,T)+\bigcirc
(e^{-<N>})  \label{s-comp2}
\end{equation}

\noindent C is an integration constant as far as the functional integration
refers but, in fact, it depends on $\rho ,T$.

The derivation is straigthforward. The origin of the compressibility term $%
S^{(c)}$ is quite clear and eqs. (38-41) of Ref. \cite{jah1} are the
equations to look at. Equation (38) is our eq. (\ref{comp-ap}) and in eqs.
(39-41) we see that each one of them has, amongst other terms, the integral $%
\int \prod n_1g_pd\{{\bf p}\}$. When the GSA for $g_p$ is used (eq. (\ref
{delf})), the integral decomposes into a sum of two integrals $\int \prod
n_1\Gamma _p\left[ 1+\Delta _p\right] d\{{\bf p}\}$. The term without $%
\Delta _p$ contributes to the ring approximation plus neglected terms (such as those
shown in Ref. \cite{jah2}) and the term with $\Delta _p$ contributes to
 $S^{(c)}$.

As $\Gamma _p$ can also be written as

\[
\Gamma _p(\{n_p\})=\frac{\prod_{\{{\bf p-1}\}\subset \{{\bf p}%
\}}n_{p-1}\prod_{\{{\bf p-3}\}\subset \{{\bf p}\}}n_{p-3}\ldots }{\prod_{\{%
{\bf p-2}\}\subset \{{\bf p}\}}n_{p-2}\prod_{\{{\bf p-4}\}\subset \{{\bf p}%
\}}n_{p-4}\ldots }
\]

\noindent using eqs. (\ref{w-def}) and (\ref{delp}) we conclude that

\begin{equation}
\frac{S^{(c)}}k=\sum_{p\geq 2}\frac 1{p!}\int d\{{\bf p}\}\left[ n_p-\frac{%
\prod_{\{{\bf p-1}\}\subset \{{\bf p}\}}n_{p-1}\prod_{\{{\bf p-3}\}\subset \{%
{\bf p}\}}n_{p-3}\ldots }{\prod_{\{{\bf p-2}\}\subset \{{\bf p}%
\}}n_{p-2}\prod_{\{{\bf p-4}\}\subset \{{\bf p}\}}n_{p-4}\ldots }\right]
\label{s-compc}
\end{equation}

\noindent It is somewhat clear that each one of these integrals is related
to p-bodies density fluctuations but a clearer explanation is to be found in
the next section. This explains the origin of naming this contribution as
compressibility related. Written in this way it is straightforward to show
that the functional derivatives are

\begin{equation}
\frac{\delta S^{(c)}/k}{\delta n_1({\bf x})}=\sum_{p\geq 1}(-1)^p\frac{%
\left\langle N\right\rangle ^p}{p!}\left[ 1+\bigcirc \left( \frac 1{%
\left\langle N\right\rangle }\right) \right] =-1+\bigcirc \left(
e^{-\left\langle N\right\rangle }\right)   \label{dsn-1}
\end{equation}

\begin{equation}
\frac{\delta S^{(c)}/k}{\delta n_s(\{{\bf x}_s\})}=\frac 1{s!}\sum_{p\geq
0}(-1)^p\frac{\left\langle N\right\rangle ^p}{p!}\left[ 1+\bigcirc \left(
\frac 1{\left\langle N\right\rangle }\right) \right] =\bigcirc \left(
e^{-\left\langle N\right\rangle }\right)   \label{dsn-s}
\end{equation}

\noindent and we arrive to eq. (\ref{ds-com}).

This result shows that, in the thermodynamic limit, the compressibility term
does not contribute to any set of equations we may derive by functional
diferentiation of a functional that includes the entropy; it only
contributes to the constraint of fixed density. Therefore, if we are after a
set of equations which are the consequence of a variational principle, we
can rightly put the compressibility term aside from all the others, while if
we are after a numerical approximation to the entropy we can assume, on
physical grounds, absolute convergence of the whole entropy series and feel
free to mix the compressibility term with all the others if numerical
convergence is improved.

\section{SERIES SUMMATION}

For the sake of this numerical goal we will cut the GSA to third order; in
this way $g_p$ can be written in two equivalent forms

\begin{equation}
g_p(\{{\bf p}\})=\left\{
\begin{array}{c}
1+\sum_{\{{\bf 2}\}\subseteq \{{\bf p}\}}h_2(\{{\bf 2}\})+\sum_{\{{\bf 3}%
\}\subseteq \{{\bf p}\}}h_3(\{{\bf 3}\})+\ldots h_p(\{{\bf p}\}) \\
\prod_{\{{\bf 3}\}\subseteq \{{\bf p}\}}\left[ 1+\Delta _3(\{{\bf 3}%
\})\right] \prod_{\{{\bf 2}\}\subseteq \{{\bf p}\}}\left[ 1+h_2(\{{\bf 2}%
\})\right]
\end{array}
\right.   \label{gp}
\end{equation}

In eq. (\ref{s-compc}) for the compressibility contribution we will sum to
infinite order three subsets of terms. These subsets are clearly identified
in the $p=3$ summand of eq. (\ref{s-compc}), i.e.

\begin{eqnarray}
\frac{S_3^{(c)}}k &=&\frac 1{3!}\int d\{{\bf 3}\}\prod_{i=1}^3n_1({\bf i}%
)\left[ g_3(\{{\bf 3}\})-\prod_{\{{\bf 2}\}\subseteq \{{\bf 3}\}}[1+h_2(\{%
{\bf 2}\})\right] =  \label{sc3} \\
&&\ \frac 1{3!}\int d\{{\bf 3}\}\prod_{i=1}^3n_1({\bf i})\left[ h_3(\{{\bf 3}%
\})-\sum_{i=1}^3\prod_{k\neq i}h_2({\bf ik})-h_2({\bf 12})h_2({\bf 13})h_2(%
{\bf 23})\right]   \nonumber
\end{eqnarray}

A) The first subset includes the contribution of the integrals $\int h_pd\{%
{\bf p}\}, p \geq 2$. The series is

\begin{equation}
\frac{S_a^{(c)}}k=\sum_{p=2}^\infty \frac 1{p!}\int d\{{\bf p}%
\}\prod_{i=1}^pn_1({\bf i})h_p(\{{\bf p}\})=\sum_{p=2}^\infty \frac{%
\left\langle C_p\right\rangle }{p!}  \label{sp-a1}
\end{equation}

The moment-cumulant relation \cite{jah1} is

\begin{equation}
C_M(\{{\bf M}\})=h_M(\{{\bf M}\})\prod_{i=1}^Mn_1({\bf i})=\sum_{k=1}^M\left%
\{ k\{{\bf m}_i\}_{{\bf M}}\right\} (-1)^{k-1}(k-1)!\prod_{i=1}^kn_{m_i}(\{%
{\bf m}_i\})  \label{sp-a2}
\end{equation}

\noindent Here, the partition of the coordinate set $\{{\bf M}\}$ in $k$
disjoint subsets $\{{\bf m}_i\}_{{\bf M}},1\leq i\leq k$ is symbolized by $%
\left\{ k\{{\bf m}_i\}_{{\bf M}}\right\} $ and therefore $%
\sum_{k=1}^M\left\{ k\{{\bf m}_i\}_{{\bf M}}\right\} $ indicates the sum
over all the partitions in $k$ subsets and for each $k$ is $1\leq i\leq k$.
In this way $\left\langle C_p\right\rangle $ is related to the integrals $%
\left\langle n_{p_i}\right\rangle =\int dp_in_{p_i}$. On the other hand, 
$\left\langle n_{p_i}\right\rangle$ can be expanded in terms of $\left\langle N^k\right\rangle $ \cite{abs}

\begin{equation}
\left\langle n_{p}\right\rangle =\left\langle N(N-1)\ldots
(N-p+1)\right\rangle =\sum_{k=1}^{p}s(p,k)\left\langle N^{k}\right\rangle
\label{sp-a3}
\end{equation}

\noindent for $p\geq 1$, where $s(p,k)$ are the Stirling numbers of first
kind. One of its definitions is that $(-1)^{p-k}s(p,k)$ is the number of
permutations of p elements which contain exactly k cycles. They satisfy the
recurrence relation

\[
s(p+1,k)=s(p,k-1)-ps(p,k),1\leq k\leq p
\]

\noindent with starting values

\[
s(p,0)=s(0,k)=\delta _{0n}
\]

We also define the r-bodies density fluctuations $\alpha _{r}$ by

\begin{equation}
\alpha _r=\left\{
\begin{array}{cc}
1 & r=1 \\
\frac{\left\langle \left( N-<N>\right) ^r\right\rangle }{<N>} & r>1
\end{array}
\right.  \label{sp-a4}
\end{equation}

The first few $\left\langle C_p\right\rangle $ are then expressed in terms
of the $\alpha _r$ and Stirling numbers as

\begin{eqnarray}
\left\langle C_2\right\rangle  &=&\left\langle N\right\rangle \left(
-1+\alpha _2\right) =\left\langle N\right\rangle \sum_{i=1}^2s(2,i)\alpha _i
\label{sp-5a} \\
\left\langle C_3\right\rangle  &=&\left\langle N\right\rangle \left(
2-3\alpha _2+\alpha _3\right) =\left\langle N\right\rangle
\sum_{i=1}^3s(3,i)\alpha _i  \label{sp-5b} \\
\left\langle C_4\right\rangle  &=&\left\langle N\right\rangle \left(
-6+11\alpha _2-6\alpha _3+\alpha _4\right) +3\left\langle N\right\rangle
^2\alpha _2^2=\left\langle N\right\rangle \sum_{i=1}^4s(4,i)\alpha
_i+3\left\langle N\right\rangle ^2\alpha _2^2  \label{sp-5c}
\end{eqnarray}

\noindent The $3\left\langle N\right\rangle ^2\alpha _2^2$ term and similar ones
from higher order $\left\langle C_p\right\rangle $ will be included in the
next partial sum. In order to sum to infinite order the contribution of each
r-bodies density fluctuations we need the result \cite{abs}

\[
\sum_{t=k}^\infty \frac{s(t,k)}{k!}x^k=\frac{\left[ \ln (1+x)\right] ^t}{t!}
\]

\noindent Therefore, the $\alpha _r$ contribution is

\begin{equation}
\Gamma _{\alpha _r}=\left\{
\begin{tabular}{ll}
$\ln 2-1$ & $r=1$ \\
$\left( \ln 2\right) ^r/r!$ & $r\geq 2$%
\end{tabular}
\right.   \label{sp-a6}
\end{equation}

\noindent and our first partial sum, which includes contributions to
infinite order of all the $\alpha _r$, is

\begin{equation}
\frac{S_a^{(c)}}k=\sum_{p=2}^\infty \frac{\left\langle C_p\right\rangle }{p!}%
=\left\langle N\right\rangle \left\{ \ln 2-1+\sum_{p=2}^\infty \frac{\left(
\ln 2\right) ^p}{p!}\alpha _p\right\}   \label{sp-a7}
\end{equation}

Let us remark on some characteristics of this result: i) $\ln 2-1$ is the
contribution in the absence of density fluctuations and it is also its high
density limit ($\alpha _p\ll 1$), ii) as $\alpha _p\rightarrow 1$ when $\rho
\rightarrow 0$, eq. (\ref{sp-a7}) vanishes in the low density limit, iii)
the series is rapidly convergent. Therefore, this sum goes in the right
direction to improve on the incompressible approximation, both in its
numerical results and in its theoretical foundation. This analysis makes
also clear why this contribution is referred to as compressibility related.
Lastly, the $\alpha _p$ are easily expressed as integrals of the correlation
functions; the first ones are

\begin{eqnarray*}
\alpha _2 &=&1+\frac 1{\left\langle N\right\rangle }\int d\{{\bf 2}\}\prod
n_1({\bf i})h_2(\{{\bf 2}\}) \\
\alpha _3 &=&-2+3\alpha _2+\frac 1{\left\langle N\right\rangle }\int d\{{\bf %
3}\}\prod n_1({\bf i})h_3(\{{\bf 3}\})
\end{eqnarray*}

B) When in the p-th term of eq. (\ref{s-compc}) the expansions given in eqs.
(\ref{gp}) are inserted, each $h_k,k<p$ expanded in terms of $h_2,\Delta _3$
and terms like the $3\left\langle N\right\rangle ^2\alpha _2^2$ which were
left aside in the first partial series included, then all the unconnected
(in the graph theory sense) terms cancel out and the first two sets of
connected diagrams are those depicted in eq. (\ref{sc3}). We first evaluate
the sum of ''star ''products of $h_2$ bonds

\begin{equation}
\frac{S_b^{(c)}}k=\sum_{p=3}^\infty \Psi _p=-\sum_{p=3}^\infty \frac 1{p!}%
\sum_{i=1}^p\int d\{{\bf i}\}n_1({\bf i})\prod_{k\neq i}\int d\{{\bf k}\}n_1(%
{\bf k})h_2({\bf ik})  \label{sp-b1}
\end{equation}

\noindent Each summand is easily evaluated as

\[
\Psi _p=\frac{\left\langle N\right\rangle }{(p-1)!}\left( -1+\alpha
_2\right) ^{p-1}
\]

\noindent and the second partial sum is

\begin{equation}
\frac{S_b^{(c)}}k=-\left\langle N\right\rangle \left[ \exp (\alpha
_2-1)-\alpha _2\right]  \label{sp-b2}
\end{equation}

\noindent Its low and high density limits are $0$ and $-e^{-1}$ respectively.

C) This series is a sum of rings very similar to eq. (\ref{s-ring}), its
first term is given in eq. (\ref{sc3}) and, as the symmetry number of
p-rings is $2p$, it can be written as

\begin{equation}
\frac{S_c^{(c)}}k=-\frac 12\sum_{p\geq 3}\frac 1p\int d\{{\bf p}%
\}\prod_{i=1}^pn_1({\bf i})h_2({\bf 12})h_2({\bf 23})\ldots h_2({\bf p1})
\label{sp-c1}
\end{equation}

\noindent which can be summed for homogeneous systems in the same way as the
original ring approximation was \cite{jah2}

\begin{equation}
\frac{S_c^{(c)}}k=-\frac{\left\langle N\right\rangle }{2\rho }\int d{\bf k}%
\left\{ \ln \left( 1-\rho \tilde h_2(k)\right) -\rho \tilde h_2(k)-\frac{%
\left( \rho \tilde h_2(k)\right) ^2}2\right\}  \label{sp-c2}
\end{equation}

\noindent where $\tilde h_2(k)=\int d{\bf r}h(r)\exp (2\pi i{\bf k}.{\bf r)}$
is the Fourier transform of $h_2(r)$ and the integration is over the ${\bf k}%
-$ space. This contribution can be added to the original ring approximation
giving a renormalized ring approximation $\tilde S^{(r)}$, which is

\begin{equation}
\frac{\tilde S^{(r)}}k=-\frac 12\sum_{p\geq 2}\frac 1p\int d\{{\bf 2p}%
\}\prod_{i=1}^{2p}n_1({\bf i})h_2({\bf 12})h_2({\bf 23})\ldots h_2({\bf 2p1})
\label{sp-c3}
\end{equation}

\noindent a sum over all even order rings and, for homogeneous systems, the
result is

\begin{equation}
\frac{\tilde S^{(r)}}k=\frac{\left\langle N\right\rangle }{2\rho }\int d{\bf %
k}\left\{ \ln \left( 1-(\rho \tilde h_2(k))^2\right) +\left( \rho \tilde h%
_2(k)\right) ^2\right\}  \label{sring-r}
\end{equation}

\section{CONCLUSIONS}

Collecting together the different results obtained, i.e. eqs. (\ref{sp-a7}),
(\ref{sp-b2}) and (\ref{sring-r}) with eqs. (\ref{s-id}) and (\ref{s-ln}) we
arrive to a new entropy equation which includes a partial summation of the
compressibility related contribution, i.e.

\begin{eqnarray}
\frac Sk &=&\left\langle N\right\rangle \left[ \frac 52-\ln (\rho \lambda
^3)\right] -\sum_{p\geq 1}\int d\{{\bf p}\}n_p(\{{\bf p}\})\omega _p(\{{\bf p%
}\})+  \label{snew} \\
&&+\frac{\left\langle N\right\rangle }{2\rho }\int d{\bf k}\left\{ \ln
\left( 1-(\rho \tilde h_2(k))^2\right) +(\rho \tilde h_2(k))^2\right\} +
\nonumber \\
&&+\left\langle N\right\rangle \left\{ \ln 2-1+\sum_{p=2}^\infty \frac{%
\left( \ln 2\right) ^p}{p!}\alpha _p\right\} -\left\langle N\right\rangle
\left[ \exp (\alpha _2-1)-\alpha _2\right]   \nonumber
\end{eqnarray}

If, for just a moment, we neglect the renormalization in the ring
approximation, the contribution due to the r-bodies density fluctuations
gives not only a theoretical understanding on the nature of the
incompressible approximation, but also a description that is essentially
correct in the low and high density limits ($0$ and $\ln 2-1-e^{-1}$
respectively). As the ring approximation grows
quite steeply when the density increases \cite{lhay1,lhay2}, its
renormalization should have the right asymptotic behaviour. Let us also
mention that these results extends trivially to mixtures (see, e.g. 
\cite{lhay2}) and, in this case, it is more convenient to work with the 
entropy per unite volume.

We have also shown that, in the thermodynamic limit, the compressibility
term only depends (as a functional) on the one body distribution function.
Therefore, this functional dependence is such that it only enters in the
constant density constraint and, in this way, the conceptual structure of
the equation for the entropy is significantly simplified. As our results 
apply to the full entropy functional, they are valuable to any functional
minimization as, e.g., those in \cite{jah2,bush}. It can also be mentioned
 that this theorem does not conflict with Laird-Haymet \cite{lhay1}%
. They obtained the correct Debye-Huckel low density expression for the
entropy by including the $S^{(id)},S^{(\ln )},$ $S^{(r)}$ plus the
compressibility related contribution of eq. \ref{comp-ap}). As this term
and, in fact, all the sums we did, vanish when $\rho \rightarrow 0$
(including the one that renormalizes the ring approximation), there is no
contradiction between ours and theirs results. Lastly, as this result does
not depend on the potential, it is also valid for the associative
Wertheim-Ornstein-Zernike equation \cite{msw1,msw2} as well as
systems with directional forces \cite{bldeve}.

This new functional provides a robust and systematic way to develop fully
analytical theories of liquids \cite{bluve1,bluve2,blub1}, which will be
examined in future work.\\

\section{ ACKNOWLEDGMENTS}

We acknowledge support from the National Science Foundation through grants
CHE-95-13558, Epscor OSR-94-52893, by the DOE-EPSCoR grant DE-FCO2-91ER75674
and CONICET grant PIP 0859/98.\\

\bibliographystyle{plain}
\bibliography{}

\end{document}